\documentclass[
 reprint,
 superscriptaddress,
 amsmath,amssymb,
 aps,
 prb,
]{revtex4-1}

\usepackage{graphicx}
\usepackage{dcolumn}
\usepackage{bm}
\usepackage[version=4]{mhchem}
\usepackage{siunitx}
\usepackage{threeparttable}
\usepackage[flushleft]{caption2}
\usepackage[colorlinks]{hyperref}
\usepackage{xcolor, soul}
\usepackage{array}

\begin{document}

\renewcommand{\figurename}{Fig.}
\renewcommand{\figureautorefname}{Fig.}
\renewcommand{\equationautorefname}{Eq.}
\newcolumntype{C}[1]{>{\centering\let\newline\\\arraybackslash\hspace{0pt}}m{#1}}

\title{Grown-in beryllium diffusion in indium gallium arsenide: An \emph{ab initio}, continuum theory and kinetic Monte Carlo study}

\author{Wenyuan \surname{Liu}}
\affiliation{Division of Physics and Applied Physics, School of Physical and Mathematical Sciences,\\ Nanyang Technological University, 21 Nanyang Link, Singapore 637371}

\author{Mahasin Alam Sk}
\author{Sergei Manzhos}
 \email{mpemanzh@nus.edu.sg}
\affiliation{%
 Department of Mechanical Engineering, National University of Singapore, Singapore 117576}

\author{Ignacio \surname{Martin-Bragado}}
 \email{nacho@synopsys.com}
\affiliation{IMDEA Materials Institute, Madrid, Spain 28906}

\author{Francis Benistant}
\affiliation{GLOBALFOUNDRIES Singapore Pte Ltd., 60 Woodlands Industrial Park D Street 2, Singapore 738406}

\author{Siew Ann \surname{Cheong}}%
\affiliation{Division of Physics and Applied Physics, School of Physical and Mathematical Sciences,\\ Nanyang Technological University, 21 Nanyang Link, Singapore 637371}

\date{\today}

\begin{abstract}
A roadblock in utilizing InGaAs for scaled-down electronic devices is its anomalous dopant diffusion behavior; specifically, existing models are not able to explain available experimental data on beryllium diffusion consistently.
In this paper, we propose a comprehensive model, taking self-interstitial migration and Be interaction with Ga and In into account.
Density functional theory (DFT) calculations are first used to calculate the energy parameters and charge states of possible diffusion mechanisms.
Based on the DFT results, continuum modeling and kinetic Monte Carlo simulations are then performed.
The model is able to reproduce experimental Be concentration profiles.
Our results suggest that the Frank-Turnbull mechanism is not likely, instead, kick-out reactions are the dominant mechanism.
Due to a large reaction energy difference, the Ga interstitial and the In interstitial play different roles in the kick-out reactions, contrary to what is usually assumed.
The DFT calculations also suggest that the influence of As on Be diffusion may not be negligible.
\end{abstract}

\maketitle

\section{Introduction} \label{sec:level1}
\par
There is significant research interest in InGaAs as a promising candidate for future generation CMOS (complementary metal-oxide semiconductor) devices (specifically, for very advanced technologies e.g. \SI{5}{\nm} node and below) due to its considerably higher electron mobility compared to Silicon.
The high electron mobility and a lattice constant that matches with InP make \ce{In_{0.53}Ga_{0.47}As} an ideal candidate for such devices.\cite{DelAlamo2011} 
Beryllium has been considered to be an important and attractive p-type dopant due to a high activation ratio and the existence of well-developed and controllable doping methods.\cite{Marcon2003} 
However, Be diffusion in InGaAs is extremely fast, with a diffusivity five orders of magnitude larger than in GaAs at the same temperature.\cite{Hu1995}
Therefore, much effort has been devoted to investigate and understand the Be diffusion behavior, including experiments and simulations.
Nevertheless, many questions are still unresolved.
In particular, there is still no agreement on the mechanism and particles' charge states for Be diffusion in InGaAs.\cite{Koumetz1995, Marcon1999, Koumetz2014}

\par
The mechanism that governs Be diffusion was widely assumed to be the kick-out mechanism (\ce{Be_{III} + I_{III} <=> Be_i}). \cite{Marcon1999, Marcon2003}
\ce{Be_{III}} denotes a Be atom in a group-III sublattice position that is considered to be immobile while \ce{Be_i} is a Be atom in an interstitial position that has a high mobility; \ce{I_{III}} represents Ga and In interstitials.
Usually, local thermodynamic equilibrium of diffusion process is also assumed.
With this model assumption, some works were able to match experimental data.
However, further clarifications about these works need to be made:
(i) samples grown under similar conditions have quite different diffusion parameters in these models;
(ii) although the local thermodynamic equilibrium assumption makes the model simple and tractable, it is not a very reasonable one, considering that Be diffusion is very fast;
(iii) the parameters used in these models, such as the reaction energy and diffusion parameters, are not extracted from or validated by other independent self-diffusion and in-diffusion experiments or \emph{ab initio} calculations, but are merely fitting parameters, making these models less predictive; and 
(iv) these models treated In and Ga as effectively the same kind of atom and assumed that As is not involved in Be diffusion.
Such assumptions are not intuitive.

\par
Some modified models have been proposed in order to overcome the above limitations.\cite{Koumetz2014}
Recently, Koumetz \emph{et al.} proposed a combined diffusion mechanism, which removes the local thermodynamic equilibrium assumption and also takes the Frank-Turnbull (dissociative) mechanism (\ce{Be_i + V_{III} <=> Be_{III}}) into account, in which \ce{V_{III}} represents Ga and In vacancies, to explain the experimental data.
Their simulation results suggested that the temperature dependence of group-III self-interstitial and of group-III vacancy effective diffusion coefficients are different.
Specifically, for temperatures above \SI{780}{\celsius}, group-III interstitial diffusion dominates the group-III vacancy diffusion, while below \SI{780}{\celsius} the situation is reversed.
Even though this model dispenses with the local thermodynamical equilibrium assumption and is more realistic than previous models that only considered the kick-out mechanism, it remains a purely phenomenological model: the choice of the mechanism and parameters is only based on fitting.

\par
A number of authors have also proposed different charge states for the diffusion mechanism:
specifically, the charge states of +1, +2 for Ga and In self-interstitials, 0, $-1$ for Ga and In vacancies, 0, +1 for Be interstitial and $-1$ for Be substitutional were proposed \cite{Koumetz1995, Koumetz1996, Marcon1997, Marcon1998a, Marcon1999, Ketata1999, Ketata1999a, Koumetz2000a, Koumetz2000b, Ihaddadene2001, Ihaddadene2002a, Koumetz2003a, Koumetz2014}.
These charge states are either inferred from works on GaAs or chosen to fit experiments.
Some of these proposed charge states are also counter-intuitive.
For example, it is known that p dopants in group VI semiconductors such as Mg interstitials in Si are positively charged, and thus a similar charge state should be expected for Be\cite{Legrain2015}.
This discrepancy provides the motivation for an independent \emph{ab initio} determination of the charge state of Be in the InGaAs system, which we will describe in \autoref{sec:level3_3}.
A realistic model should explain experimental data obtained under different annealing conditions in one consistent way and based on as few assumptions as possible.
To the best of our knowledge, such a model is still lacking.

\par
In this work, a comprehensive and physically-based model of diffusion behavior of grown-in Be in InGaAs is presented.
To select possible mechanisms for the Be diffusion, we calculated the reaction energies and diffusion barriers of a variety of possible mechanisms using density functional theory.
The results suggest that the energies required for the Frank-Turnbull mechanism are much higher than for the kick-out mechanism, and so in the temperature range relevant to experiments, the Frank-Turnbull mechanism can be safely ignored.
Furthermore, among the kick-out reactions, the energies required for reactions involving Ga and In are quite different.
Contrary to the previous models, the roles of Ga and In in Be diffusion are different, and these elements ideally should not be lumped together.
The influence of As on Be diffusion may not be negligible since the reaction energies for As being kicked out by Be are comparable with Ga/In being kicked out.
We then build a diffusion model which based on reaction energies and diffusion barriers calculated from first principles.
By implementing this model in the Object Kinetic Monte Carlo simulator (OKMC) MMonCa,\cite{Martin-Bragado2013} we are able to reproduce experimental data under different annealing temperatures and durations in a consistent way. 

\section{Methodology} \label{sec:level2}
\subsection{Density functional theory} \label{sec:level2_1}
\par
The calculations were performed using density functional theory \cite{Koch2001, Parr1980} with the generalized gradient approximation and the Perdew-Burke-Eznerhof functional (GGA-PBE)\cite{Perdew1996} as implemented in Vienna \emph{ab initio} simulation package (VASP).\cite{Kresse1993, Kresse1996, Kresse1996a}
The projector augmented wave method (PAW)\cite{Blochl1994, Kresse1999} was used to describe the interaction between the atomic cores and electrons.
The valence configurations of the atoms were: Arsenic (As) $4s^24p^3$, Gallium (Ga) $4s^24p^1$, Indium (In) $5s^25p^1$ and Beryllium (Be) $2s^22p^0$.
A \num{2x2x2} Monkhorst-Pack\cite{Pack1977} k-point mesh and a cutoff of \SI{400}{\eV} were used for structure optimization.
Atomic positions and cell vectors, where applicable, were relaxed using the conjugate gradient (CG) algorithm until all force components were less than \SI{0.01}{\eV\per\angstrom}.
The single-point energy calculations on the structures, relaxed using the cutoff of \SI{400}{\eV} and \num{2x2x2} k-points, were performed with \num{6x6x6} k-points to achieve converged defect formation energies and diffusion barriers.
A tetrahedron method with Bl\"{o}chl\cite{Blochl1994} corrections was used for the partial occupancies. The density-of-state (DOS) calculations were performed using \num{5x5x5} k-points meshes.
The energy barrier for Be, Ga and In-atom diffusion were calculated using the climbing-image nudged elastic band (CI-NEB) method;\cite{Henkelman2000, Henkelman2000a} the force tolerance in the CI-NEB calculations was \SI{0.05}{\eV\per\angstrom}.

\par
To ensure that the simulation cell is of size amenable to the calculations, we used the stoichiometry \ce{In_{0.5}Ga_{0.5}As_1}(abbreviated in the following as InGaAs), as was done in previous works.\cite{Komsa2012b, Komsa2012c}
The simulation cell size of about \SI[product-units = single]{12.0 x 12.0 x 11.9}{\angstrom} was used and is sufficient to neglect inter-cell interactions of the defects.
The geometries of doped-InGaAs are fully relaxed.
The CI-NEB calculations are performed under fixed cell.
The charges on atoms of pure and doped InGaAs crystal are calculated using Bader analysis\cite{Tang2009}.

\subsection{Kinetic Monte Carlo simulation} \label{sec:level2_2}
\par
The Be diffusion process was modeled by object kinetic Monte Carlo using the MMonCa code\cite{Martin-Bragado2013}.
In the framework of MMonCa, the diffusion process is composed of succeeding events, either reaction or migration, which occur at different rates. 
In this study, we consider only kick-out reactions, surface trappings or injections, migrations and transitions between charge states.
Once reactants approach each other within the reaction distance, the reaction will occur with the probability
\begin{equation}\label{eq:reaction rate}
  P = \exp\left(-\frac{E_{re}}{k_BT}\right)
\end{equation}
when $E_{re}$ is positive, otherwise the probability is 1.
$E_{re}$ is the reaction barrier, $k_B$ is the Boltzmann constant and $T$ is the temperature.
In this study, such values are calculated according to the transition state theory.
The interface between air and the InGaAs sample is the sink and the source of all self-interstitials: \ce{Ga_i}, \ce{In_i}; the trapping rate and injection rate are also determined by \autoref{eq:reaction rate}

\par
Self-interstitials and the Be interstitial can migrate in random directions with a fixed migration distance $\lambda$, while substitutional Be atoms are assumed to be immobile.
The use of a fixed migration distance is justified by the crystal structure of InGaAs where elementary diffusion steps have all similar lengths.
The migration rate is computed as
\begin{equation}\label{eq:migration rate}
  \nu_m = \nu_m^0\exp\left(-\frac{E_m}{k_BT}\right),
\end{equation}
where $\nu_m^0$ is the migration attempt frequency; $E_m$ is the migration barrier.

\par
Point defects in semiconductors usually have various electronic states.
If we assume that a point defect X can be in three different charge states for example, singly negative, neutral, singly positive, we will denote this point defect as $X^j$, with $j=-, 0, +$.
Then the relative concentrations are
\begin{equation}\label{eq:charge states transformation}
  \frac{[X^j]}{[X^{j+1}]} = \frac{g^j}{g^{j+1}}\exp\left(\frac{e_F-e\left(j+1, j\right)}{k_BT}\right),
\end{equation}
where $g^j$ stands for the degeneracy factor, $e_F$ is the Fermi level and $e(j+1, j)$ represents the energy level associated to the charge transition between $X^j$ and $X^{j+1}$.
In the MMonCa framework, a point defect can transform between its different charge states.
The transition rate is determined by \autoref{eq:charge states transformation}.
In fact, charge state transformation is much faster than the diffusion process, so the relative concentration of charge states is almost in equilibrium everywhere.

\par
So far we have considered only thermal Brownian motion.
In our case, doping will introduce an electric field, which will introduce an additional driving force into the diffusion equation.
We also need to consider this term.
In the MMonCa framework, the ratio between the migration frequency along the electric field and opposite to it for a point defect with charge nq is
\begin{equation}\label{eq:electric_drift}
  \frac{\nu_{m,+}}{\nu_{m,-}} = \exp\left(\frac{nq\varepsilon\lambda}{k_BT}\right),
\end{equation}
where $n$ is the charge number, $q$ is the elementary charge, $\varepsilon$ is the electric field intensity and $\lambda$ is the migration distance.

\par
Our simulation box has dimensions of \num{1202x15x15} nm with periodic boundary conditions along the $y$ and $z$ directions.
The initial Be dopant atoms are all set to the substitutional state and Ga or In interstitials are set at their equilibrium concentrations.
The Be concentration profiles are extracted after annealing durations corresponding to those used in the experiments have passed.

\section{Results and discussion} \label{sec:level3}
\subsection{Comparison between different diffusion mechanisms: DFT energetics} \label{sec:level3_1}
\par
The following types of defects and elementary migration steps in Be-doped InGaAs were considered in DFT calculations:
(a) migration of an interstitial atom (\ce{Be_i}, \ce{Ga_i} and \ce{In_i}) to a neighboring interstitial site,
(b) movement of a Be atom from a substitutional site (\ce{Be_{As}}: Be at an As site, \ce{Be_{Ga}}: Be at a Ga site or \ce{Be_{In}}: Be at an In site) to an interstitial site leaving either As, Ga or In vacancy (\ce{V_{As}}, \ce{V_{Ga}} and \ce{V_{In}})(Frank-Turnbull mechanism) and
(c) movement of a \ce{Be_i} atom to a substitutional site by displacing Ga/In/As to an interstitial site (kick-out mechanism).

\par
\begin{figure}
  \centering
  \includegraphics[width=0.5\textwidth]{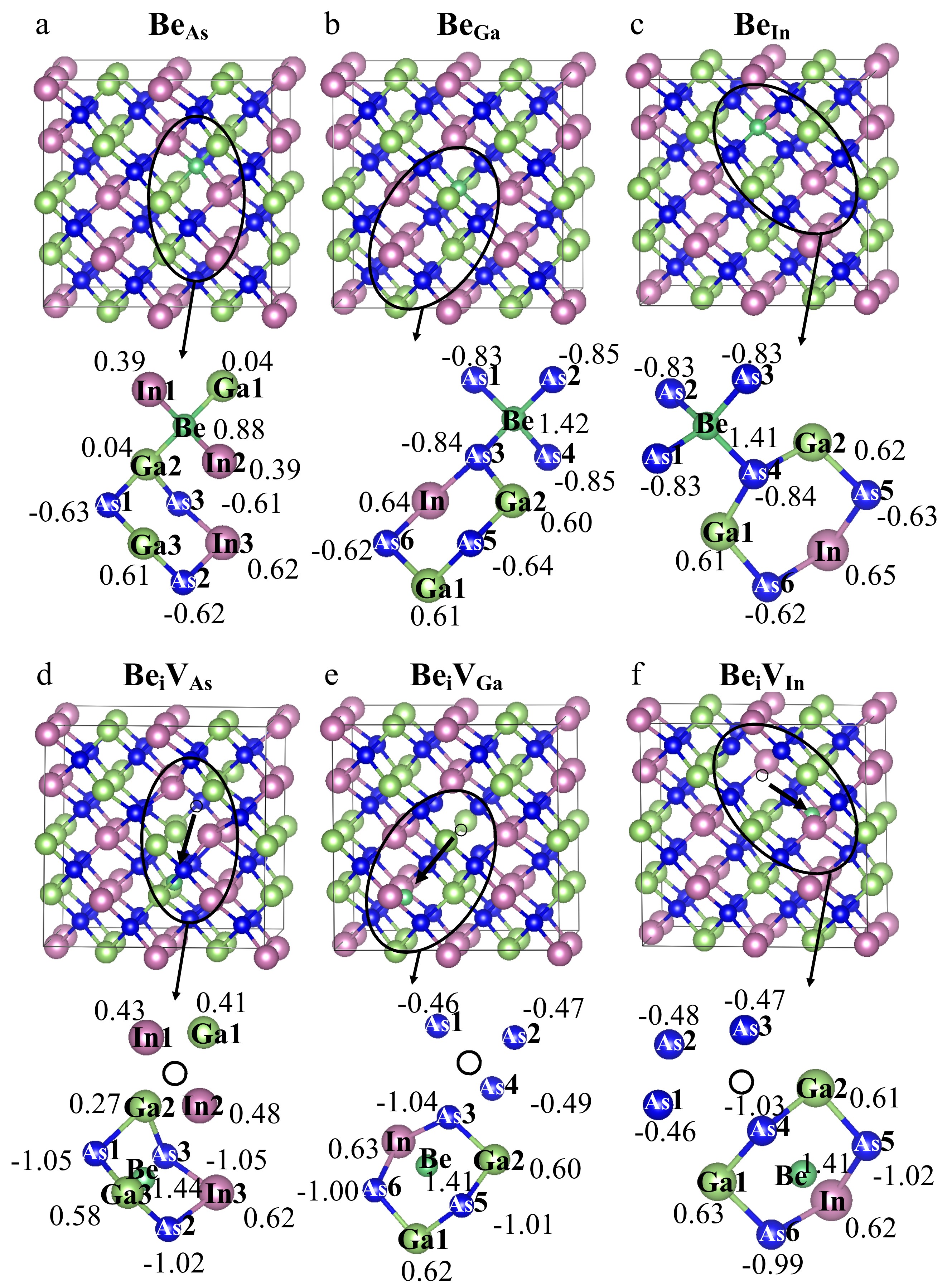}
  \caption{Structures of (a) \ce{Be_{As}}, (b) \ce{Be_{Ga}} and (c) \ce{Be_{In}} doped InGaAs. (d), (e) and (f) are the structures of doped InGaAs after the \ce{Be_{As}}, \ce{Be_{Ga}} and \ce{Be_{In}} atoms moved to an interstitial site leaving \ce{V_{As}}, \ce{V_{Ga}} and \ce{V_{In}}, respectively. Insets show the Be atom and its neighbouring atoms with Bader charges. Violet spheres \text{-} In, large green spheres \text{-} Ga, blue spheres \text{-} As and small green spheres \text{–} Be atom. Visualization here and elsewhere is by VESTA\cite{Momma2011}.}
 \label{fig:dissocative and charge state}
\end{figure}
\begin{figure}
  \centering
  \includegraphics[width=0.5\textwidth]{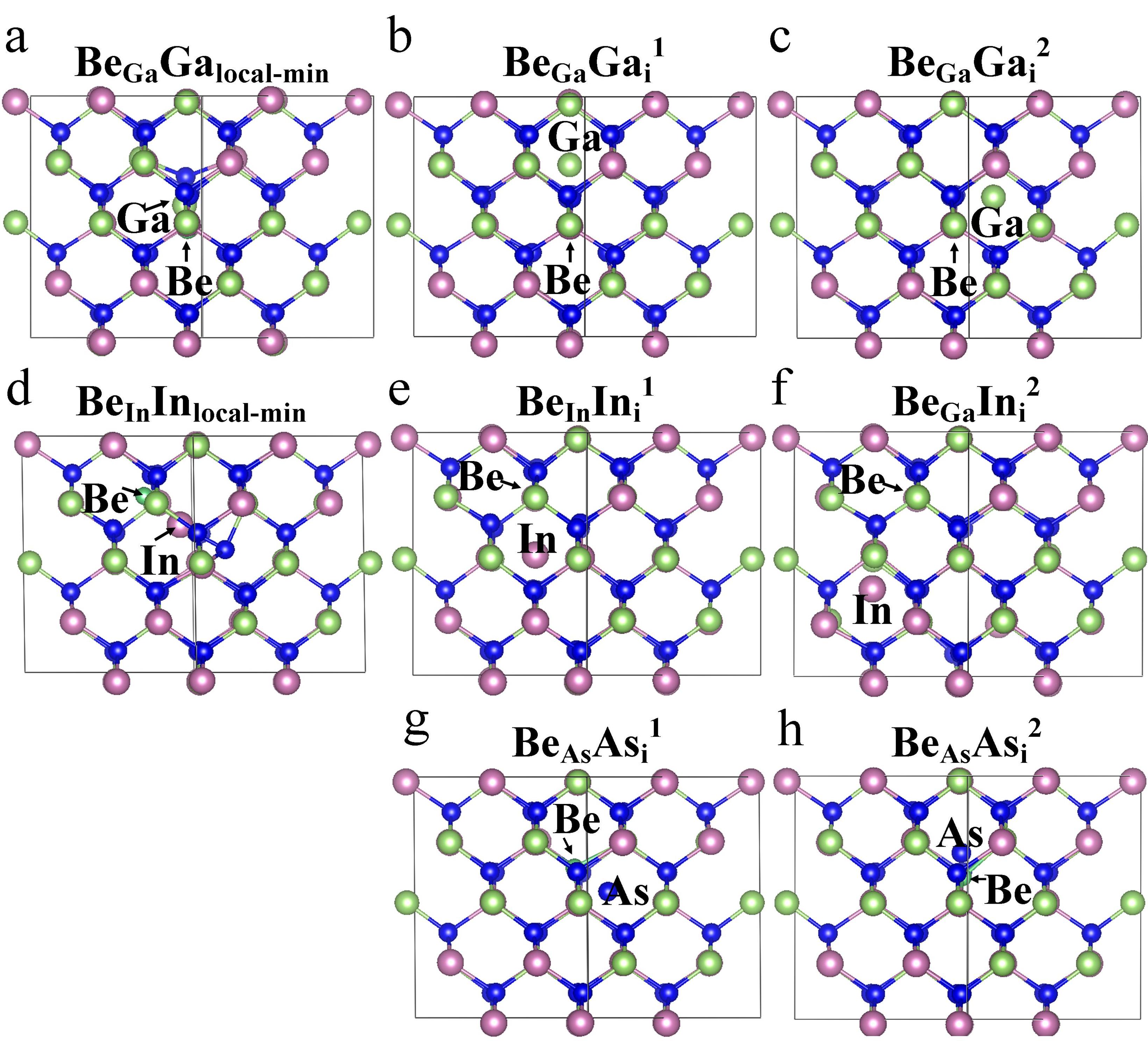}
  \caption{Structure of InGaAs with (a,b,c) a substitutional Be atom and a Ga atom at an interstitial site (local minimum, site1 and site2, respectively) and (d,e,f) a substitutional Be atom and an In atom at an interstitial site (local minimum, site1 and site2, respectively) and (g,h) a substitutional Be atom and an As atom at an interstitial site (site1 and site2, respectively)} \label{fig:kick-out mechanism}
\end{figure}

In \autoref{tab:dissociative mechanism}, we list the reaction energies for a Be atom diffusion from a \ce{Be_{As}}, \ce{Be_{Ga}} or \ce{Be_{In}} substitutional site to an interstitial site leaving either \ce{V_{As}}, \ce{V_{Ga}} or \ce{V_{In}}, respectively (see  \autoref{fig:dissocative and charge state}).
The positive reaction energies mean that the diffusion of a Be-atom from a substitutional site to an interstitial site is an endothermic reaction.
\autoref{tab:kick-out mechanism} lists reaction energies for various kick-out reactions, and corresponding configurations are shown in \autoref{fig:kick-out mechanism}.
In these reactions, the lattice Ga, In, and As atoms can be displaced by a Be atom into two kinds of interstitial sites (site1 and site 2), as shown in \autoref{fig:kick-out mechanism}.
Comparing with reaction energies needed for the kick-out mechanism (\autoref{tab:kick-out mechanism}), Frank-Turnbull mechanism requires much higher energies.
We can therefore conclude that the kick-out mechanism is the dominant mechanism, at least in the annealing temperature range (\SI{700}{\celsius}-\SI{900}{\celsius}) in experiments.

\begin{table}
  \centering
  \caption{Energy required for Be atom diffusion from a substitutional site (\ce{Be_{As}}, \ce{Be_{Ga}} or \ce{Be_{In}}) to an interstitial site (leaving \ce{V_{As}}, \ce{V_{Ga}} or \ce{V_{In}}) calculated by DFT.} \label{tab:dissociative mechanism}
  \begin{threeparttable}
  \begin{tabular}{cc}
    \hline
    Reaction & Reaction energy(eV)\tnote{1} \\
    \hline
    \ce{InGaAs-Be_{As} -> InGaAs-Be_iV_{As}} & 0.87 \\
    \ce{InGaAs-Be_{Ga} -> InGaAs-Be_iV_{Ga}} & 2.50 \\
    \ce{InGaAs-Be_{In} -> InGaAs-Be_iV_{In}} & 2.54 \\
    \hline
  \end{tabular}
  \begin{tablenotes}
  \item[1] Reaction energy = $E$(\ce{InGaAs-Be_iV_M}) - $E$(\ce{InGaAs-Be_M}), where M = As, Ga and In
  \end{tablenotes}
  \end{threeparttable}
\end{table}

\par
We also note from \autoref{tab:kick-out mechanism} that In and Ga appear to have different roles.
For simplification, previous models all treated In and Ga as the same idealized group-III element.
However, as can be seen from \autoref{tab:kick-out mechanism}, their properties are quite different:
the kick-out of Ga by a Be interstitial is exothermic while the kick-out of In by a Be interstitial is endothermic.
It is worth noting that the reaction energies of kick-out of As by a Be interstitial is comparable with kick-out of the III-group elements, which is ignored by previous models.
However the role of As is complicated by the swap reaction (see \autoref{fig:dissociative and energies}e), so that full picture on how As influences Be diffusion is difficult to work out.

\begin{table}
  \centering
  \caption{Reaction energy required for an interstitial Be atom to migrate to a lattice site of InGaAs by kicking out a Ga, In or As atom to an interstitial site calculated by DFT.} \label{tab:kick-out mechanism}
  \begin{threeparttable}
  \begin{tabular}{cc}
    \hline
    Reaction & Reaction energy(eV)\tnote{1} \\
    \hline
    \ce{InGaAs + Be_i -> InGaAs-Be_{Ga} + Ga_i(site1)} & -0.63  \\
    \ce{InGaAs + Be_i -> InGaAs-Be_{Ga} + Ga_i(site2)} & -0.72  \\
    \ce{InGaAs + Be_i -> InGaAs-Be_{In} + In_i(site1)} & 0.31  \\
    \ce{InGaAs + Be_i -> InGaAs-Be_{In} + In_i(site2)} & 0.11 \\
    \ce{InGaAs + Be_i -> InGaAs-Be_{As} + As_i(site1)} & 0.37  \\
    \ce{InGaAs + Be_i -> InGaAs-Be_{As} + As_i(site2)} & 0.17 \\
    \hline
  \end{tabular}
  \begin{tablenotes}
  \item[1]Reaction energy = $E$(InGaAs-Be+\ce{M_i}) - $E$(InGaAs+\ce{Be_i}), M = Ga, In and As
  \end{tablenotes}
  \end{threeparttable}
\end{table}

\subsection{Diffusion paths and barriers: DFT kinetics} \label{sec:level3_2}
\par
\begin{figure}
  \centering
  \includegraphics[width=0.5\textwidth]{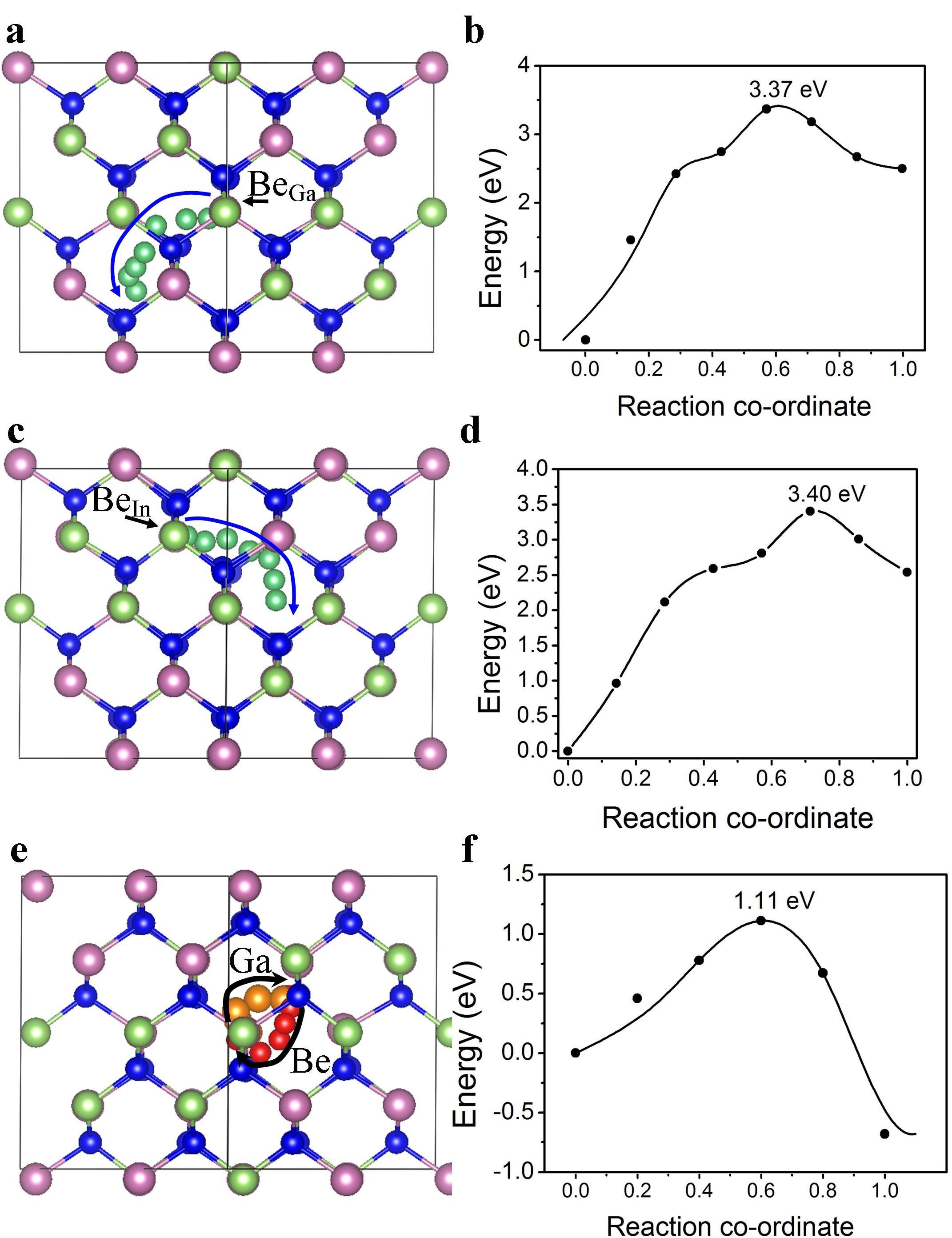}
  \caption{Diffusion path of (a) \ce{Be_{Ga}} to \ce{Be_i} leaving \ce{V_{Ga}} vacancy (b) \ce{Be_{In}} to \ce{Be_i} leaving \ce{V_{In}} vacancy and (c) \ce{Be_{As}} to \ce{Be_{Ga}} and \ce{Ga_{Ga}} to \ce{Ga_{As}} in InGaAs. (b), (d) and (f) are the calculated potential energy curve along the diffusion path for diffusion in (a), (c) and (e), respectively. Be at Ga, In and As site at equilibrium corresponds to 0.0 on the abscissa. Be after diffusion at equilibrium corresponds to 1.0. The solid line in panels (b, d and f) is given to guide the eyes.} \label{fig:dissociative and energies}
\end{figure}

The higher reaction energies calculated in \autoref{sec:level3_1} suggest that the Frank-Turnbull mechanism is not important.
The calculated diffusion barriers for the \ce{Be_{Ga}} and \ce{Be_{In}} diffusions to an interstitial site are 3.37 and \SI{3.40}{\eV} (see \autoref{fig:dissociative and energies}a, b, c and d, respectively).
The high reaction energies and barriers allow us to discard both reactions.
For the diffusion of the \ce{Be_{As}} to an interstitial site leaving \ce{V_{As}}, the path is more complex:
the \ce{Be_{As}} swaps the position with a neighbouring Ga-atom (see  \autoref{fig:dissociative and energies}e, f) instead of migrating to an interstitial site.
The reaction energy for this swapping reaction is \SI{-0.68}{\eV} and the barrier is \SI{1.11}{\eV}.
This result suggests that \ce{Be_{As}} is an unstable configuration for the Be dopant. 
Therefore there will only be transient occupation of As sites by Be at high temperatures.

\par
\begin{figure}
  \centering
  \includegraphics[width=0.5\textwidth]{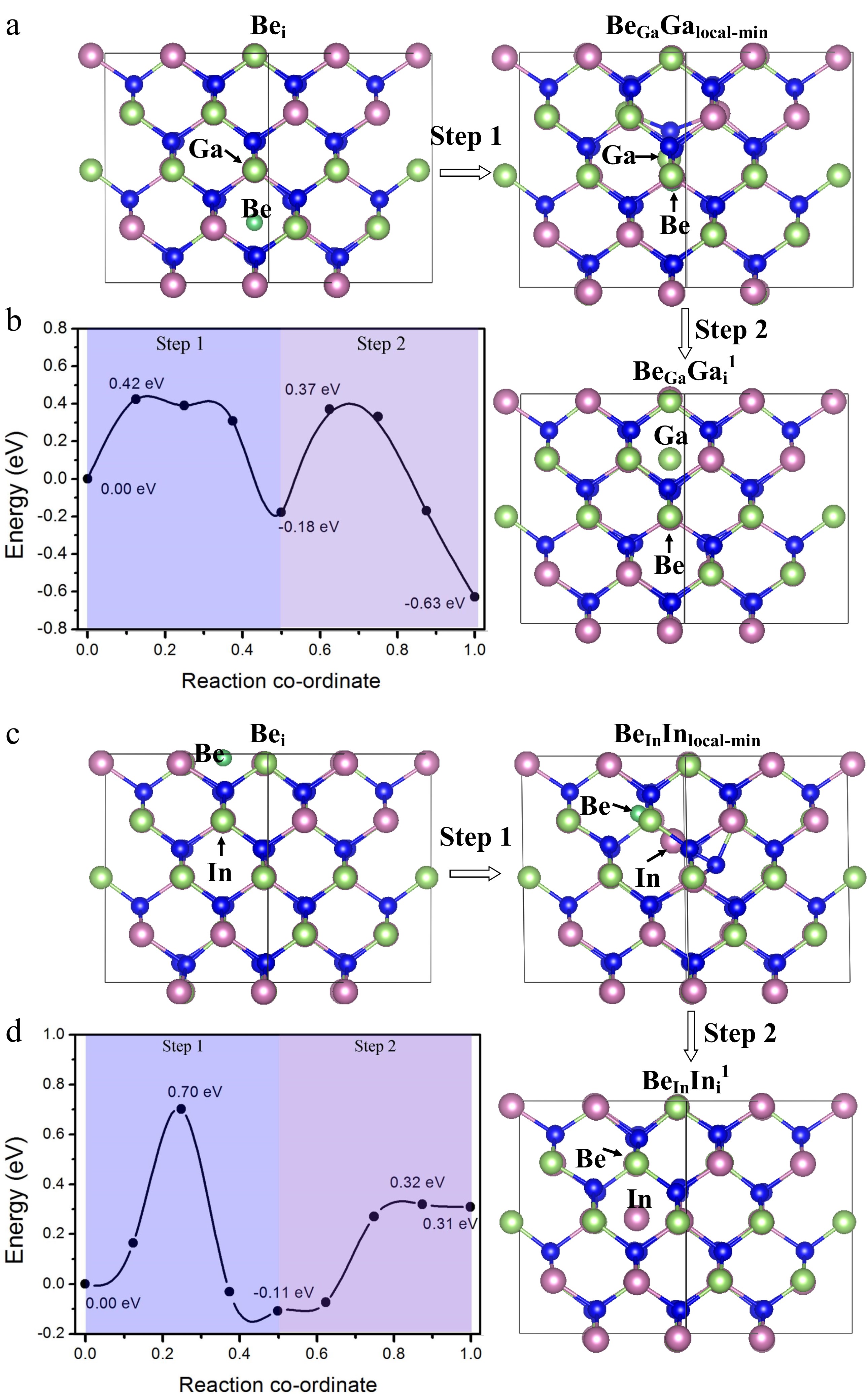}
  \caption{Steps of \ce{Be_i} atom diffusion to (a) \ce{Be_{Ga}} by kicking out the Ga atom and (c) \ce{Be_{In}} by kicking out the In atom to an interstitial site. The calculated potential energy curve along the diffusion path of the \ce{Be_i} to (b) \ce{Be_{Ga}} by kicking out the Ga atom to an interstitial site and (d) \ce{Be_{In}} by kicking out the In atom to an interstitial site. Be atom at an interstitial site (\ce{Be_i}) corresponds to 0.0 and Be at either Ga or In site (\ce{Be_{Ga}} or \ce{Ge_{In}}) with either Ga or In atom moved to an interstitial site corresponds to 1.0 on the abscissa. The solid line in panels (b and d) is given to guide the eyes.} \label{fig:kickout steps and energies}
\end{figure}
\begin{figure}
  \centering
  \includegraphics[width=0.5\textwidth]{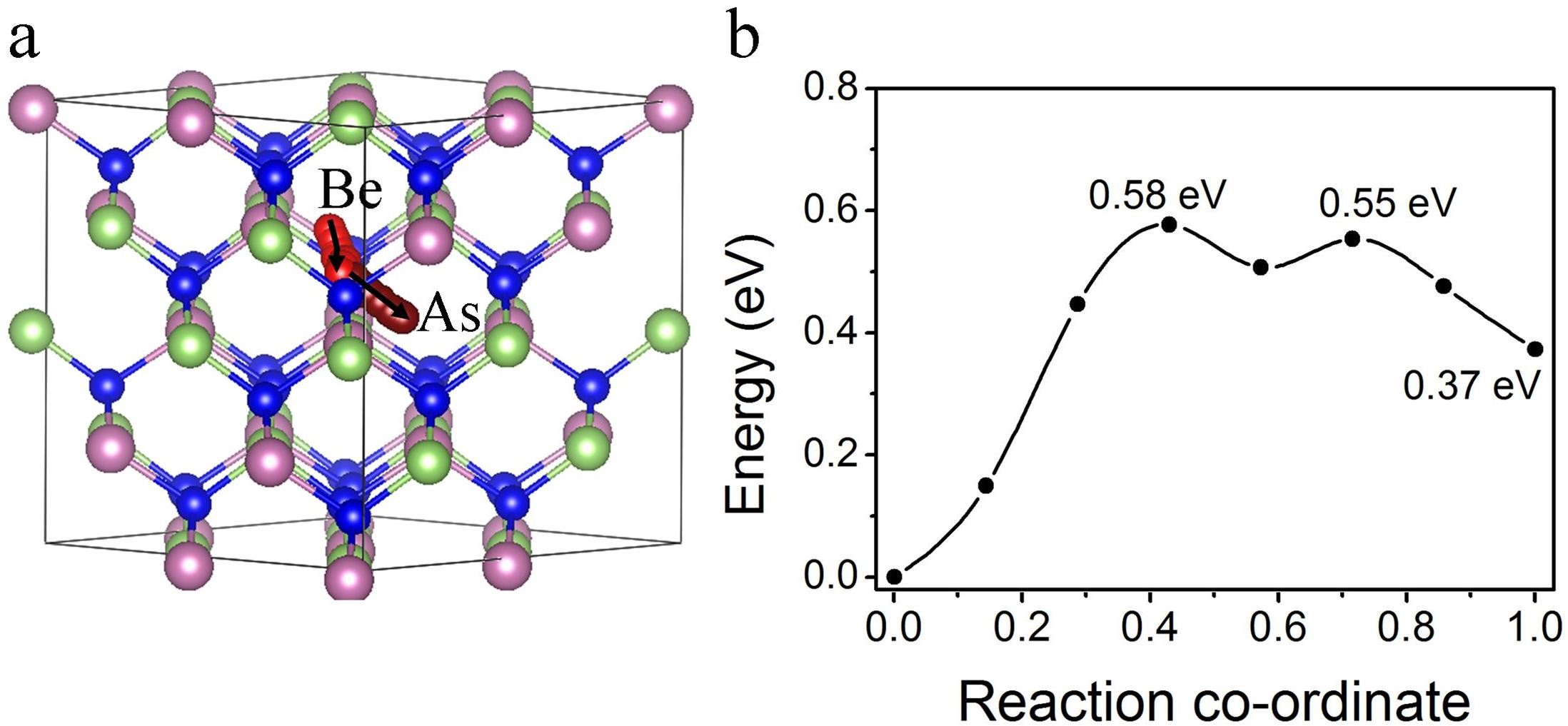}
  \caption{(a) Diffusion path for As kick-out by \ce{Be_i} in InGaAs and (b) the calculated potential energy curve along the diffusion path. The Be atom at an interstitial site corresponds to 0.0; and Be atom at As site and As atom at an interstitial site after the kick out corresponds to 1.0 on the abscissa. The solid line is given to guide the eyes.} \label{fig:BeAs kickout path}
\end{figure}

We have also calculated the diffusion paths and barriers for the kick-out mechanism.
Both the kick-out of Ga and the kick-out of In can be divided into two steps separated by a local minimum, as shown in \autoref{fig:kickout steps and energies}.
In the case of Ga being kicked out by the Be interstitial, barriers for steps 1 and 2 are 0.42 and \SI{0.55}{\eV} respectively; whereas for the In being kicked out, barriers are 0.70 and \SI{0.43}{\eV} respectively.
For the kick-out of a Ga atom, the energy barrier for step 1 is lower than for the kick-out of In.
The local energy minima suggest the existence of quasi-stable Be-Ga and Be-In split-interstitial structures or \emph{dumbbell} configurations. 
In contrast, we did not observe any local energy minimum for the As atom being kicked out (see \autoref{fig:BeAs kickout path}), and the diffusion barrier is \SI{0.58}{\eV}.

\par
It is important to note that although both In and Ga being kicked out can be divided into two steps, the physics are different:
for the kick out of Ga, the intermediate state has a higher energy than the final state and therefore will not influence the reaction outcome;
for the kick out of In, the intermediate state is more like a sink because its energy is lower than that of both the initial state and the final state.
This emphasizes the importance of treating In and Ga separately.

\par
\begin{figure}
  \centering
  \includegraphics[width=0.5\textwidth]{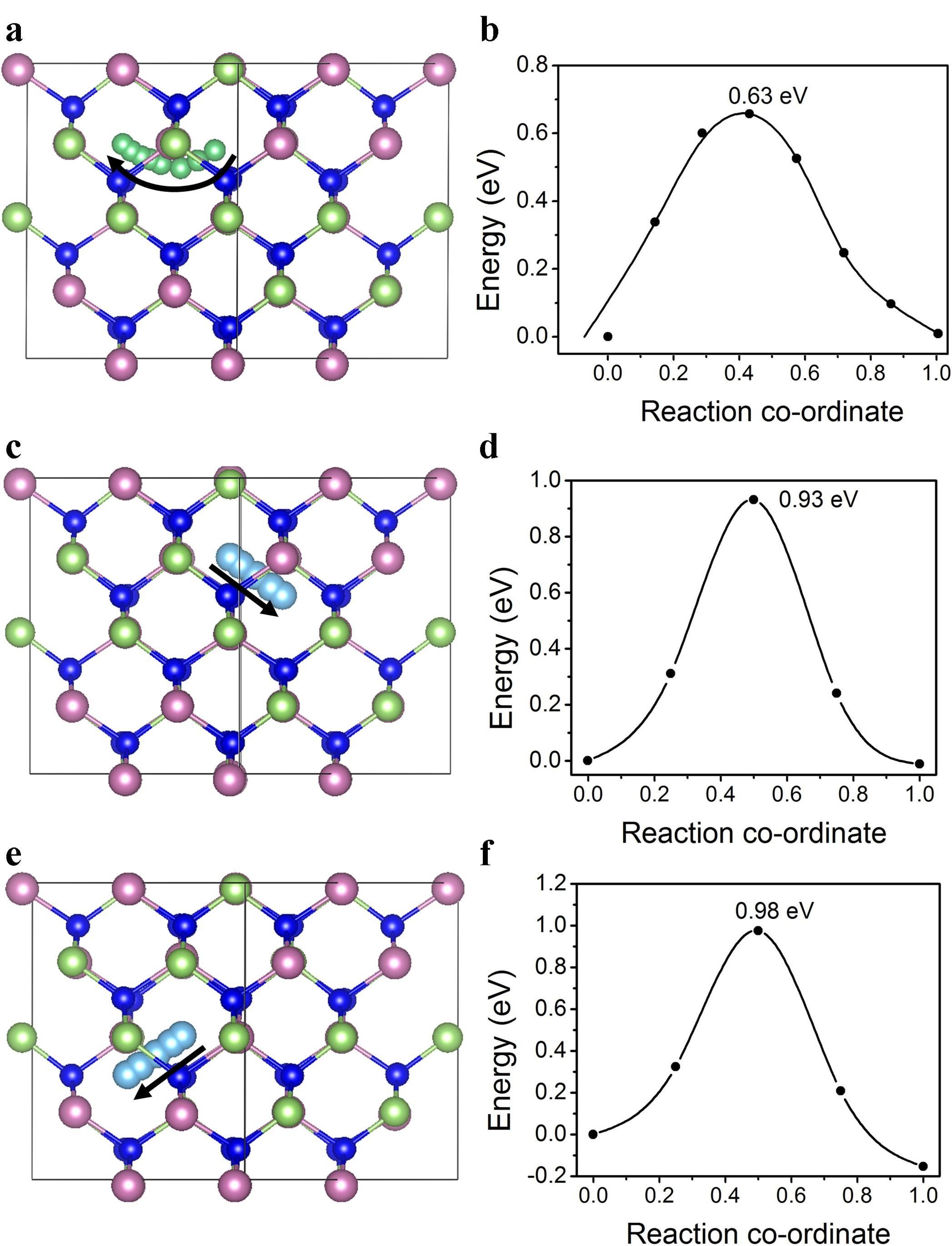}
  \caption{(a) Diffusion path of (a) \ce{Be_i}, (c) \ce{Ga_i} and (e) \ce{In_i} in InGaAs. (b), (d) and (f) are the calculated potential energy curves along the diffusion path in (a), (c) and (e), respectively. Be, Ga and In at an interstitial site at equilibrium corresponds to 0.0. Be, Ga and In at another interstitial site after diffusion at equilibrium corresponds to 1.0 on the abscissa. The solid line in panels (b, d and f) is given to guide the eyes.} \label{fig:migration energies}
\end{figure}

The calculated diffusion barriers for the migration of \ce{Be_i}, \ce{Ga_i} and \ce{In_i} from an interstitial site to a neighboring interstitial site in InGaAs are 0.63, 0.93 and \SI{0.98}{\eV}, respectively (see \autoref{fig:migration energies}).
We note that the diffusion paths of \ce{Ga_i} and \ce{In_i} are straight and symmetric whereas the diffusion path of \ce{Be_i} is curved and asymmetric.

\subsection{Charge state analysis: DFT} \label{sec:level3_3}
\par
\begin{figure}
  \centering
  \includegraphics[width=0.5\textwidth]{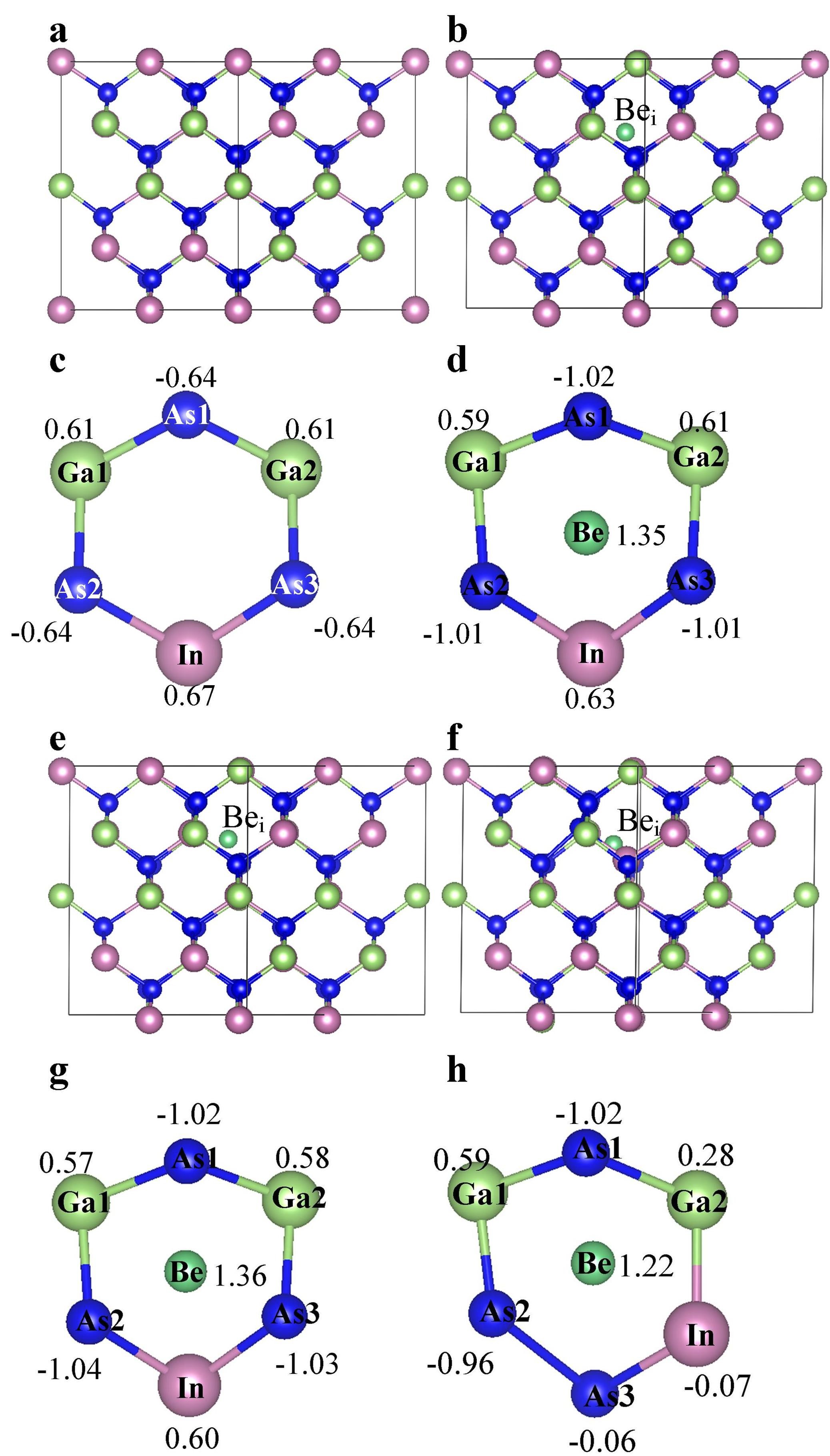}
  \caption{(a) Pure and (b) \ce{Be_i} doped InGaAs (\ce{Be_i-InGaAs}). (e) \ce{Be_i-InGaAs} with charged supercell (-1e) and (f) \ce{Be_i-InGaAs} with flipping the position of As with In atom near \ce{Be_i}. Atoms domain with Bader charges in (c) pure, (d) \ce{Be_i-InGaAs}, (g) \ce{Be_i-InGaAs} with charged supercell and (h) \ce{Be_i-InGaA} with flipping the position of In with As atom near \ce{Be_i}.} \label{fig:pure and Bei interstitial charge states}
\end{figure}

The average Bader charges on As, Ga and In atom in pure InGaAs crystal are $-0.64e$, $+0.61e$ and $+0.67e$, respectively as shown in  \autoref{fig:pure and Bei interstitial charge states}c.
There is one symmetry-unique interstitial binding site in the InGaAs crystal for Be atom insertion (\ce{Be_i}) as shown in \autoref{fig:pure and Bei interstitial charge states}b,d.
The Be atom is on top of the plane formed by three nearest As atoms and equidistant from each As atoms.
The calculated Bader charge on \ce{Be_i} atom is $+1.35e$ (see \autoref{tab:charge state} and \autoref{fig:pure and Bei interstitial charge states}d) in the equilibrium position but becomes $1.15e$ in the transition state for diffusion.
\begin{table}
  \centering
  \caption{Calculated Bader charges on Be, Ga and In atoms in doped InGaAs.} \label{tab:charge state}
  \begin{tabular}{C{0.4\columnwidth}C{0.4\columnwidth}}
  \hline
  Atom type          & Bader charge[$e$] \\
  \hline
  \ce{Be_i}          & 1.35           \\
  \ce{Ga_i}          & 0.36(0.28)      \\
  \ce{In_i}          & 0.43(0.37)      \\
  \ce{Be_{As}}       & 0.88            \\
  \ce{Be_{Ga}}       & 1.42            \\
  \ce{Be_{In}}       & 1.41            \\
  \ce{Be_iV_{As}}    & 1.44            \\
  \ce{Be_iV_{Ga}}    & 1.41            \\
  \ce{Be_iV_{In}}    & 1.41            \\
  \ce{Be_{Ga}Ga_i^1} & 1.42            \\
  \ce{Be_{Ga}Ga_i^2} & 1.40            \\
  \ce{Be_{In}In_i^1} & 1.38           \\
  \ce{Be_{In}In_i^2} & 1.41           \\
  \hline
  \end{tabular}
\end{table}
By examining the Bader charges of the neighboring As, Ga, and In atoms, we find that the inserted \ce{Be_i} atom donates electron density to its neighbors.
The Bader charges on the As atoms (see  \autoref{fig:pure and Bei interstitial charge states}c, d) significantly changes upon Be insertion, going from $-0.64e$ to $-1.01e$ suggesting that the dominant electron donation is from the Be atom to neighbouring As atoms.
There are two symmetry-unique interstitial binding sites in InGaAs for Ga (\ce{Ga_i}) and In (\ce{In_i}) atom insertion. The total energy difference between the two \ce{Ga_i} sites is \SI{0.01}{\eV} whereas for \ce{In_i} total energy difference is \SI{0.15}{\eV}.
The calculated Bader charges on \ce{Ga_i} atoms in the two sites are $0.36e$ and $0.28e$ respectively in the equilibrium position and $0.31e$ in the transition state for diffusion.
Similarly, charges on \ce{In_i} in the two sites are $0.43e$ and $0.37e$, respectively, in the equilibrium position and $0.40e$ in the transition state for diffusion.

\par
The calculated Bader charges on the \ce{Be_{As}}, \ce{Be_{Ga}} and \ce{Be_{In}} are $+0.88e$, $+1.42e$ and $+1.41e$, respectively.
The \ce{Be_{As}} atom donates electrons to Ga and In whereas \ce{Be_{Ga}} and \ce{Be_{In}} donate electron to the As atoms (see  \autoref{fig:dissocative and charge state}).
Calculated Bader charges on the Be atom for \ce{Be_iV_{As}}, \ce{Be_iV_{Ga}} and \ce{Be_iV_{In}} configurations are $+1.44e$, $+1.41e$ and $+1.41e$, respectively (see  \autoref{tab:charge state} and  \autoref{fig:dissocative and charge state}).
The total charges on the atoms bonded to the Be atom in \ce{Be_{As}}, \ce{Be_{Ga}} and \ce{Be_{In}} doped InGaAs are $+0.86e$, $-3.37e$ and $-3.33e$ respectively, which change to $+1.59e$, $-2.46e$ and $-2.46e$, respectively, when the Be atom diffuses to the interstitial site.

\par
To investigate the kick-out mechanism, we have replaced either an Ga or an In atom with a Be atom from the interstitial site and placed that Ga or In atom in interstitial sites; \ce{Be_{Ga}Ga_i^1}, \ce{Be_{Ga}Ga_i^2}, \ce{Be_{Ga}In_i^1}, \ce{Be_{Ga}In_i^2} (see \autoref{fig:kick-out mechanism}).
The calculated Bader charges on the Be in configuration \ce{Be_{Ga}Ga_i^1}, \ce{Be_{Ga}Ga_i^2}, \ce{Be_{In}In_i^1} and \ce{Be_{In}In_i^2} are $+1.42e$, $+1.40e$, $+1.38e$ and $+1.41e$, respectively (see \autoref{tab:charge state}).

\par
\begin{figure}
  \centering
  \includegraphics[width=0.5\textwidth]{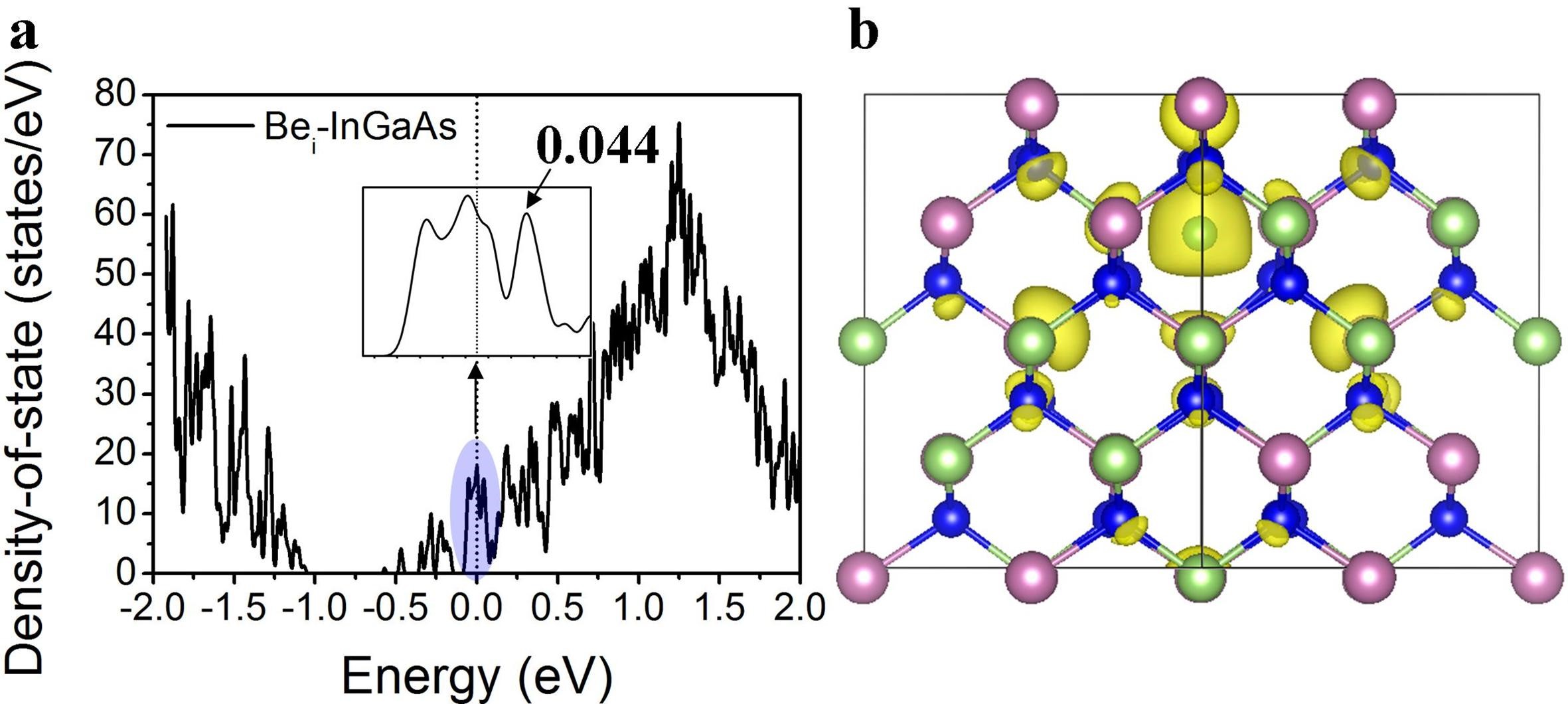}
  \caption{(a) Total DOS plot of \ce{Be_i}-doped InGaAs. The Fermi level is set to zero (dotted line). The inset is the zoomed up DOS plot where the energy axis (x-axis) ranges from -0.1 to \SI{0.1}{\eV}. (b) The partial (band decomposed) charge density plot of band at \SI{0.044}{\eV} above the Fermi level.} \label{fig:dos of Bei}
\end{figure}

Like any charge definition, the Bader charge analysis used here is but one definition of atomic charges; it is done based on the topology of the electron density and naturally results in non-integer charge states.
These charge states can, however, be assigned to integer charge states and compared to those used in the KMC model by considering the valence shell build-up of the dopant.
For \ce{Be_i}, the Bader charge state of about $1.4e$ therefore corresponds to the donation of both its $2s$ electrons to the host structure (specifically, to the conduction band, as shown in  \autoref{fig:dos of Bei} , i.e. \ce{Be_i} is a n-type dopant).
This is corroborated by the fact that the zero spin state is preserved upon Be insertion (i.e. both valence electrons of Be are in the conduction band).
\ce{Be_i} can therefore be assigned a 0-K state of $+2$.
Likewise, the \ce{Ga_i} Bader charge of about $+0.3e$ suggests that the Ga atom does not lose any of its valence electrons and the charge of $0.3e$ is due to charge redistribution following bond formation with the host atoms.
We can assign to \ce{Ga_i} a 0-K state of $0$. Similarly we assign a charge state of $0$ to \ce{In_i}.

\par
\begin{figure}
  \centering
  \includegraphics[width=0.5\textwidth]{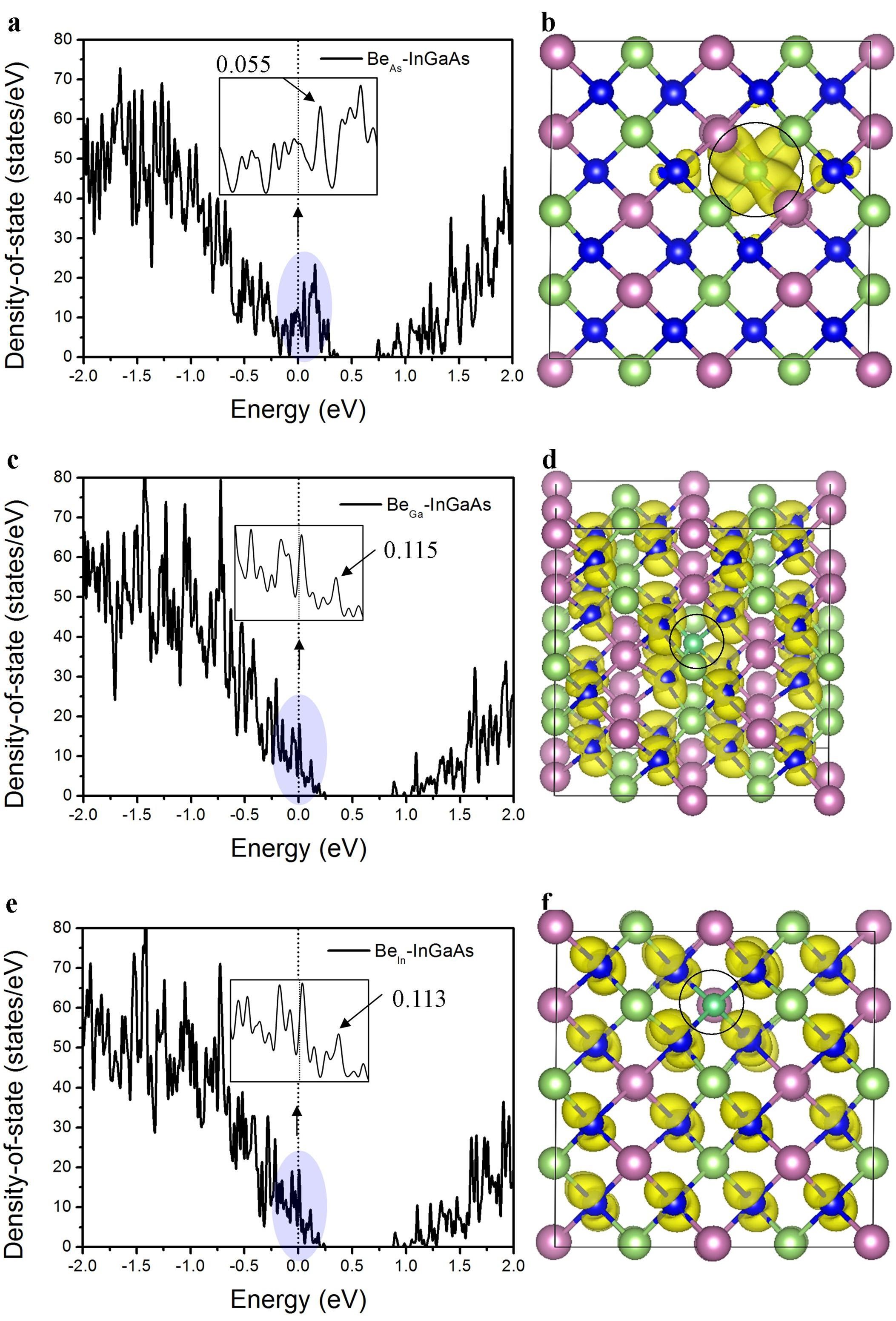}
  \caption{Total DOS plot of (a) \ce{Be_{As}}, (c) \ce{Be_{Ga}} and (e) \ce{Be_{In}} doped InGaAs. The Fermi level is set to zero (dotted line). The insets are the zoomed up DOS plots where the energy axis (x-axis) ranges from -0.2 to 0.2 eV. The partial (band decomposed) charge density plot of bands at 0.055, 0.115 and \SI{0.113}{\eV} above the Fermi level in (b) \ce{Be_{As}}, (d) \ce{Be_{Ga}} and (f) \ce{Be_{In}} doped InGaAs, respectively.}\label{fig:dos Besubstitutional}
\end{figure}

Be atoms substituting a Ga or an In atom (\ce{Be_{Ga}} and \ce{Be_{In}}) also loose both their valence electrons, except in this case the electrons occupy states in the valence band, as shown in  \autoref{fig:dos Besubstitutional}.
This means that substitutional Be atoms play the role of p dopants but remain positively charged (about $+1.4e$) in the host structure similarly to Mg doping of Si or Ge \cite{Ho1979, Legrain2016}.
We have confirmed that the positive charge does not noticeably change even when imposing a negative charge on the supercell (see \autoref{fig:pure and Bei interstitial   charge states}).
We therefore assign 0-K states of $+2$ to both \ce{Be_{Ga}} and \ce{Be_{In}}.

\par
The above charge calculation is done for an ideal ordered crystal, however the real InGaAs is a random alloy.
To test the effects due to local disorder and non-stoichiometry, we exchanged the positions of As and In atoms in the vicinity of the \ce{Be_i} atom. \autoref{fig:pure and Bei interstitial   charge states}g, h shows the resulting structures and atomic charges. This antisite defect does not significantly change the charge state of Be; however, it changes the charge states of the involved As and In atoms to near-neutral.
The total energy cost of such an exchange is \SI{1.43}{\eV}, so the effect of these configurations can be neglected.

\par
\begin{figure}
  \centering
  \includegraphics[width=0.5\textwidth]{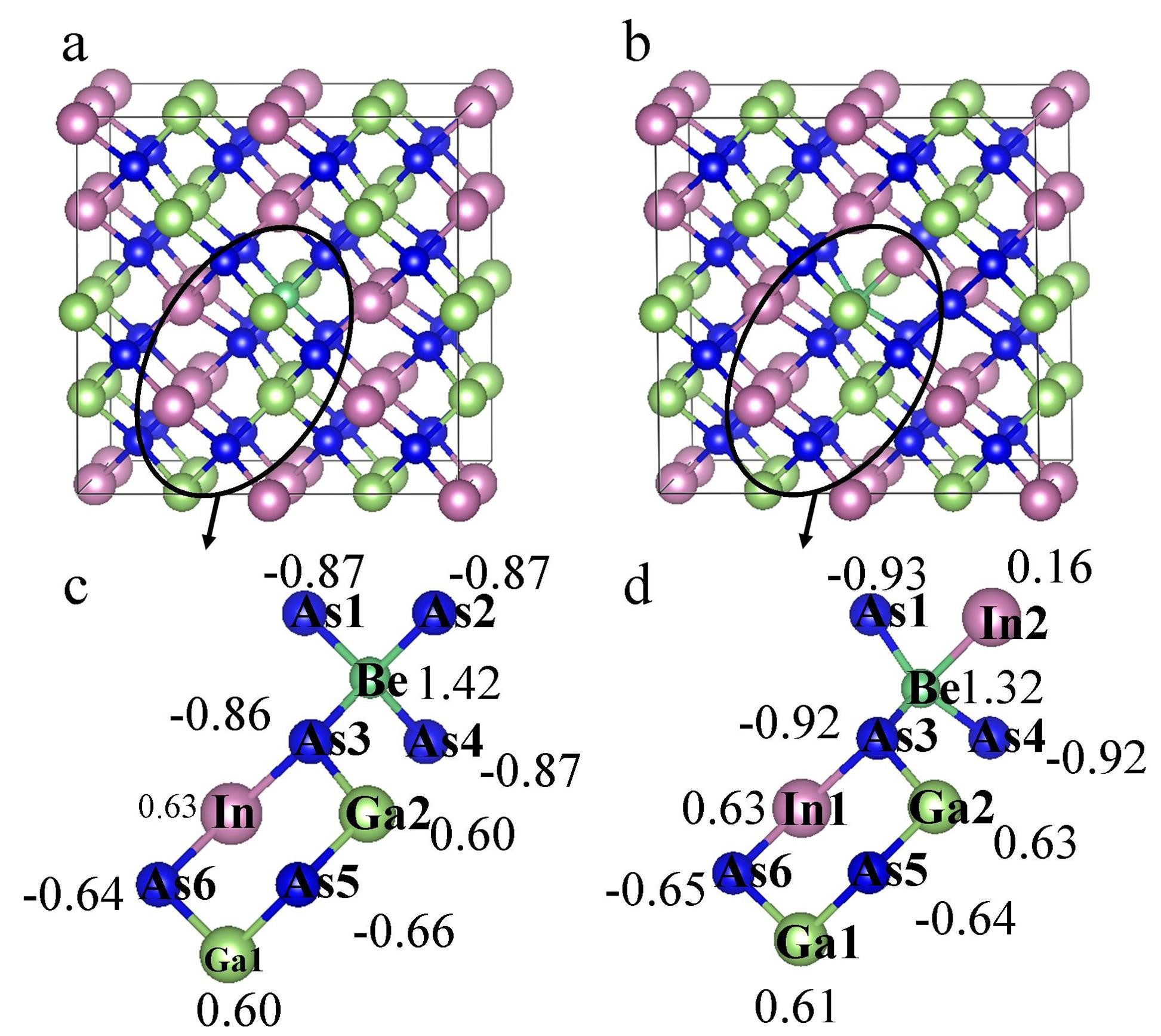}
  \caption{(a) \ce{Be_{Ga}-InGaAs} with charged supercell ($-1e$) and (b) \ce{Be_{Ga}-InGaAs} with flipping the position of As with In atom near \ce{Be_{Ga}}. Atoms surrounding the Be dopant with Bader charges in (c) \ce{Be_{Ga}-InGaAs} with charged supercell ($-1e$) and (d) \ce{Be_{Ga}-InGaAs} with flipping the position of In with As atom near \ce{Be_{Ga}}.} \label{fig:BeGa charge}
\end{figure}

Also, when we exchange the In atom with an As atom near the \ce{Be_{Ga}} site, there are significant differences of charges on the exchanged In and As atoms (see \autoref{fig:BeGa charge}c), but no significant change in the charge state of \ce{Be_{Ga}}. Similar to \ce{Be_i-InGaAs}, the flipped In and As atoms can have a zero charge state in \ce{Be_{Ga}-InGaAs}.
However, the energy required to exchange the In atom with an As atom is \SI{2.06}{\eV} and therefore the presence of such antisite defects can be neglected.
The calculations show that non-stoichiometric effects will surely influence the charge distribution, however, due to their high defect formation energy, they are not expected to have a significant effect on the model.

\par
\begin{figure}
  \centering
  \includegraphics[width=0.5\textwidth]{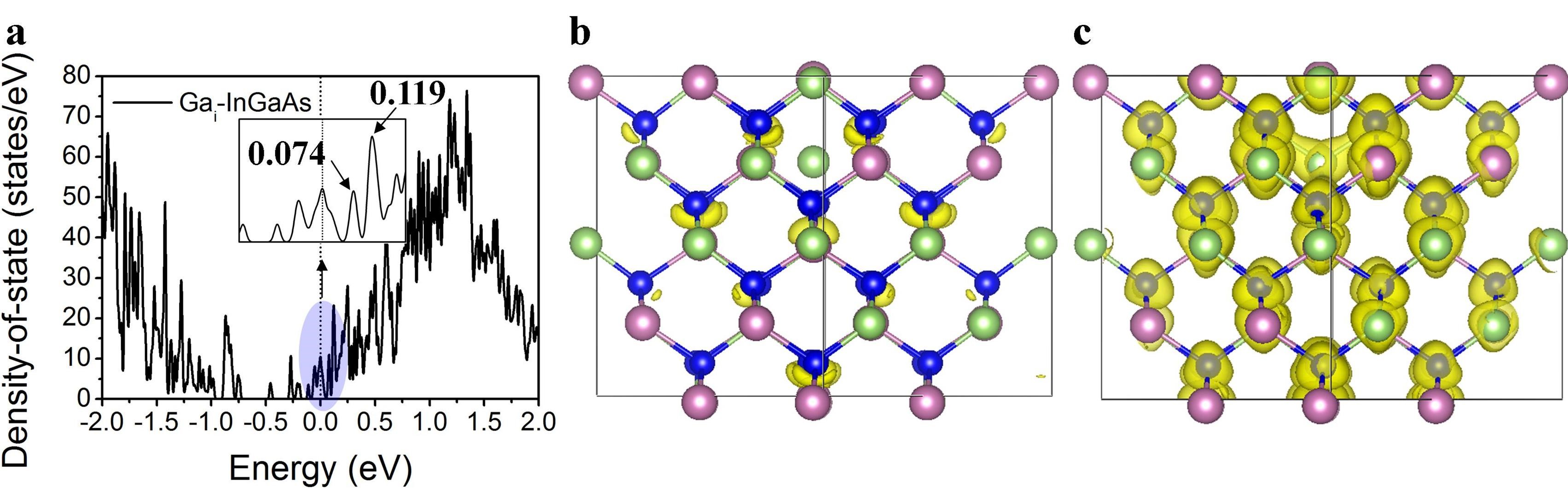}
  \caption{(a) Total DOS plot of \ce{Ga_i}-doped InGaAs. The Fermi level is set to zero (dotted line). The inset is the zoomed up DOS plot where the energy axis (x-axis) ranges from -0.2 to \SI{0.2}{\eV}. (b) and (c) The partial (band decomposed) charge density of band at 0.074 and \SI{0.119}{\eV} above the Fermi level.} \label{fig:dos of Gai}
\end{figure}

We emphasize that the above charge state assignment is valid at zero temperature (for which the \emph{ab initio} calculations are done).
At finite temperature, available empty states just above the Fermi level ($E_f$) will be occupied.
If these states are localized on the dopant, this will generate other dopant charge states.
We therefore analyse possible charge states relevant for the KMC simulations below.
For \ce{Be_i}, there is a Be-centered state \SI{0.044}{\eV} above $E_f$, as shown in \autoref{fig:dos of Bei}.
According to the Boltzmann distribution, the partial occupancy of this state is $18\%$ at \SI{300}{\kelvin}.
This suggests that at \SI{300}{\kelvin} $18\%$ of \ce{Be_i} atoms will be in state $+1$ rather than $+2$.
At temperatures relevant to experiments as well as KMC simulations, i.e. ~ \SI{1100}{\kelvin}, $63\%$ of \ce{Be_i} will be in state $+1$. Similarly, we identified a state at \SI{0.119}{\eV} above $E_f$ for \ce{Ga_i}. This state is not strictly Ga-centred but has a significant localization on the \ce{Ga_i}, as shown in \autoref{fig:dos of Gai}.
The occupation probability of this state (and therefore the fraction $\mathrm{Ga}_{\mathrm{i}}^{-}$ is 1\%) at \SI{300}{\kelvin} and $28\%$ at \SI{1100}{\kelvin}. For \ce{Be_{Ga}} and \ce{Be_{In}}, we did not find low-lying Be-centred states likely to be occupied at finite temperatures.

\subsection{Continuum calculations} \label{sec:level3_4}
\par
Continuum calculations of diffusion in solids are widely used due to their conceptual simplicity and numerical flexibility.
However, the complexity grows rapidly with the number of reactions involved.
This is the case for Be diffusion in InGaAs where multiple elementary steps contribute to Be diffusion as we have elucidated in the preceding sections.
Nevertheless, with a few additional assumptions, continuum calculations can still provide valuable insights as an exploratory tool, as we will demonstrate as follows.

\par
The continuum treatment is based on Fick's law of diffusion.
Once the mechanism is assumed, one can write down the corresponding differential equations and solve them numerically.
The equations parametrically depend on diffusivities and reaction energies, which one can fit to match experimental data.
Previous simulations have shown that the simulated Be concentration profile is not very sensitive to the effective diffusivity, but is instead very sensitive to the charge state assumed in the diffusion mechanism.
The charge state can therefore be reliably determined from the fit.\cite{Marcon1999}

\par
Our goal is to explain many experiments using a single, consistent set of parameters.
Although many experiments of Be diffusion in InGaAs are reported in literature, they are done under different experimental conditions.
Therefore, in the following, we consider only experiments where the only difference in conditions are annealing temperatures and durations.\cite{Ketata1999, Koumetz2003a, Marcon2003, Koumetz2014}
Specifically, we consider experiments where:
(i) the doping process is done by gas source molecular beam epitaxy rather than ion implantation or in-diffusion from an extended source, both of which will induce implantation damage or surface effect, which are difficult to account for in a simulation;
(ii) the gas source molecular beam epitaxy parameters are similar.
It is well known that such parameters will strongly influence the material's transport properties.

\par
\begin{figure}
  \centering
  \includegraphics[width=0.5\textwidth]{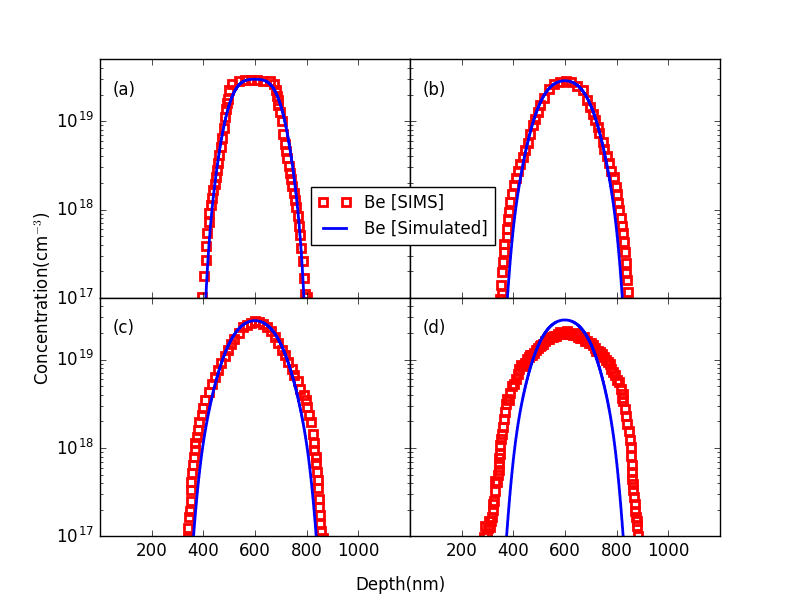}
  \caption{Experimental and simulated Be profiles for the mechanism of \autoref{eq:kick out v1} obtained with the continuum method under different annealing conditions: (a)\SI{180}{\second} at \SI{700}{\celsius}, (b)\SI{120}{\second} at \SI{750}{\celsius}, (c)\SI{60}{\second} at \SI{800}{\celsius}, (d)\SI{10}{\second} at \SI{900}{\celsius}.\cite{Marcon2003, Koumetz2003a, Ketata1999}} \label{fig:kick out v1}
\end{figure}

We first assumed the following trial diffusion mechanism:
\begin{equation}\label{eq:kick out v1}
  \mathrm{Be}_{\mathrm{III}}^{++} + \mathrm{III}_{\mathrm{i}}^0 \leftrightharpoons \mathrm{Be}_{\mathrm{i}}^{++},
\end{equation}
where $\mathrm{Be}_{\mathrm{III}}^{++}$ denotes a doubly positive Be atom in a group-III sublattice position, $\mathrm{III}_{\mathrm{i}}^0$ represents a neutral Ga or In interstitial, and $\mathrm{Be}_{\mathrm{i}}^{++}$ is a doubly positive Be atom in an interstitial position.
The charge states are selected based on DFT estimates of state occupancies at experimental temperatures (see \autoref{sec:level3_3}).
The corresponding diffusion equations are the following:
\begin{equation}\label{eq:diffusion equation}
\begin{aligned}
  \frac{\partial C_i}{\partial t} = \frac{\partial}{\partial t}\left(D_i\frac{\partial C_i}{\partial x} - nD_i\frac{C_i}{p}\frac{\partial p}{\partial x} \right) - \frac{\partial C_s}{\partial t}, \\
  \frac{\partial C_I}{\partial t} = \frac{\partial}{\partial t}\left(D_I\frac{\partial C_I}{\partial x} - rD_I\frac{C_I}{p}\frac{\partial p}{\partial x}\right) + \frac{\partial C_s}{\partial t},
\end{aligned}
\end{equation}
where $C_i$ is the concentration of $\mathrm{Be}_{\mathrm{i}}^{n+}$ where $n=2$, $C_I$ is the concentration of $\mathrm{I}_{\mathrm{III}}^{r+}$ where $r=0$, $C_s$ is the concentration of $\mathrm{Be}_{\mathrm{III}}$, $D_i$ and $D_I$ are the diffusivities of $\mathrm{Be}_{\mathrm{i}}$ and $\mathrm{III}_{\mathrm{i}}^{r+}$; $p$ is the hole concentration; $n$ and $r$ are Be interstitial and III-group interstitial charge numbers respectively.
The best fit achieved with  \autoref{eq:kick out v1} at several temperatures is shown in \autoref{fig:kick out v1}. We note that this model fit the data reasonably well at \SI{700}{\celsius}, \SI{750}{\celsius}, \SI{800}{\celsius}, but shows a serious discrepancy at \SI{900}{\celsius}.

\par
\begin{figure}
  \centering
  \includegraphics[width=0.5\textwidth]{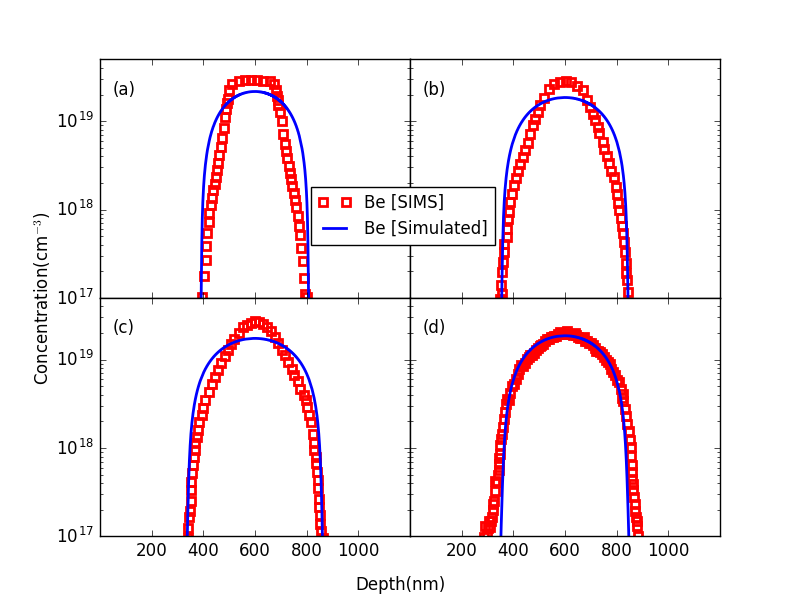}
  \caption{Experimental and simulated Be profiles for the mechanism of \autoref{eq:kick out v2} obtained with the continuum method under different annealing conditions: (a)\SI{180}{\second} at \SI{700}{\celsius}, (b)\SI{120}{\second} at \SI{750}{\celsius}, (c)\SI{60}{\second} at \SI{800}{\celsius}\cite, (d)\SI{10}{\second} at \SI{900}{\celsius}\cite{Marcon2003, Koumetz2003a, Ketata1999}} \label{fig:kick out v2}
\end{figure}
\begin{figure}
  \centering
  \includegraphics[width=0.5\textwidth]{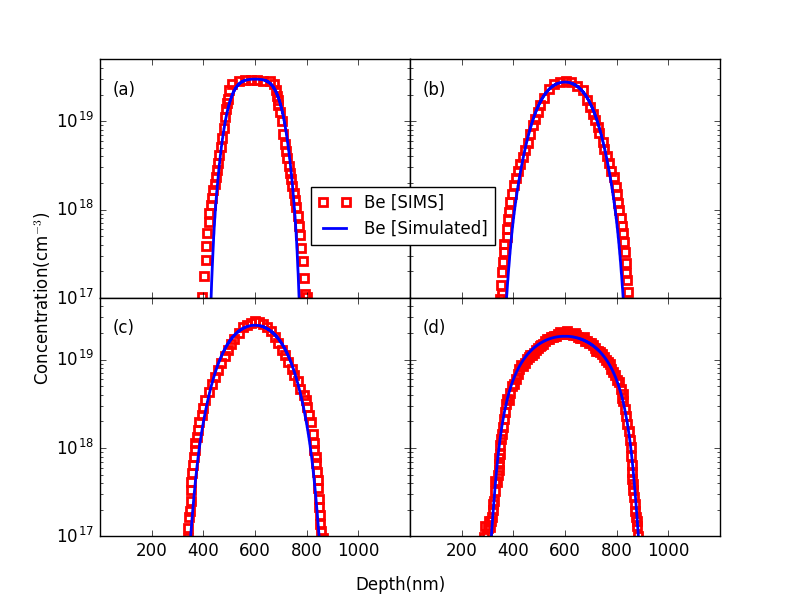}
  \caption{Experimental and simulated Be profiles for the mechanism of \autoref{eq:kick out v3} obtained with the continuum method under different annealing conditions: (a)\SI{180}{\second} at \SI{700}{\celsius}, (b)\SI{120}{\second} at \SI{750}{\celsius}, (c)\SI{60}{\second} at \SI{800}{\celsius}, (d)\SI{10}{\second} at \SI{900}{\celsius}.\cite{Marcon2003, Koumetz2003a, Ketata1999}} \label{fig:kick out v3}
\end{figure}

Therefore we tried another diffusion mechanism:
\begin{equation}\label{eq:kick out v2}
  \mathrm{Be}_{\mathrm{III}}^{++} + \mathrm{III}_{\mathrm{i}}^- \rightleftharpoons \mathrm{Be}_{\mathrm{i}}^+,
\end{equation}
where $\mathrm{III}_{\mathrm{i}}^-$ represents a singly negative Ga or In interstitial and $\mathrm{Be}_{\mathrm{i}}^{+}$ is a singly positive Be atom in an interstitial position.
The corresponding partial differential equations are similar to \autoref{eq:diffusion equation} but with $n=1$ and $r=-1$.
The best fit achieved by \autoref{eq:kick out v2} for different temperatures is shown in \autoref{fig:kick out v2}.
We see that it agrees with the data at \SI{900}{\celsius}, but does not fit well at \SI{700}{\celsius}, \SI{750}{\celsius}, \SI{800}{\celsius}, especially in the high concentration region.
This is the opposite of  \autoref{eq:kick out v1}.
Since we have already shown based on DFT calculations that at low temperatures, $\mathrm{Ga}_{\mathrm{i}}^0$ population dominates, and $\mathrm{Ga}_{\mathrm{i}}^-$ will only have a significant population at the highest temperatures among considered here, that the models assuming a single $\mathrm{Ga}_{\mathrm{i}}$ charge state only fit the experimental data in different temperature ranges is consistent with DFT results.
This suggests that we should include at least two mechanisms with different temperature dependences, one dominating at lower temperatures, another at higher temperatures.
The integrated model proposed is:
\begin{equation} \label{eq:kick out v3}
\begin{aligned}
  \mathrm{Be}_{\mathrm{III}}^{++} + &\mathrm{III}_{\mathrm{i}}^0 \rightleftharpoons &\mathrm{Be}_{\mathrm{i}}^{++}, \\
              &\Updownarrow              &\Updownarrow  \\
  \mathrm{Be}_{\mathrm{III}}^{++} + &\mathrm{III}_{\mathrm{i}}^- \rightleftharpoons &\mathrm{Be}_{\mathrm{i}}^+.
\end{aligned}
\end{equation}
The horizontal reactions are elementary diffusion reactions, while the vertical arrows indicate changes in charge state.
Such changes are due to changes in electronic state occupancies and are much faster than diffusion reactions; therefore, we can assume that the equilibrium between different charge states is instantaneous and this equilibrium is determined by local Fermi level and transition energy. The results of fits using this model are shown in  \autoref{fig:kick out v3}.
Clearly, the model performs very well over the entire temperature range.

\subsection{Kinetic Monte Carlo simulations} \label{sec:level3_5}
\par
Although the continuum treatment can provide a good fit to experimental data, its application is limited to simple diffusion mechanisms.
The reason is that the number of differential equations will grow as the diffusion mechanism becomes more and more complex, the differential equations need to be coupled and advanced numerical methods need to be used.
Compared with the continuum treatment, kinetic Monte Carlo (KMC) simulation is very suitable for simulating very complex diffusion mechanisms because one simply needs to add reactions into an event list, and the algorithm will handle complicated correlations implicitly.

\par
\begin{figure}
  \centering
  \includegraphics[width=0.5\textwidth]{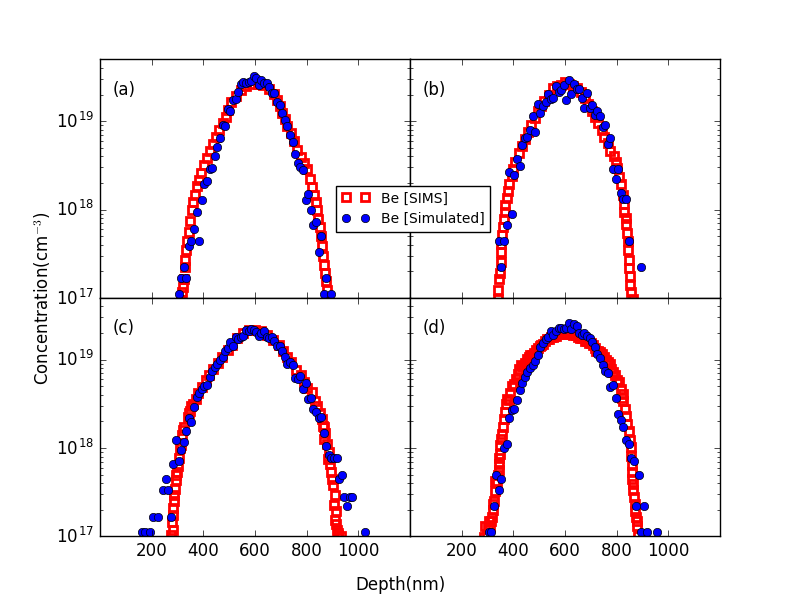}
  \caption{Experimental and simulated Be profiles for mechanism \autoref{eq:kmc mechanism} by kinetic Monte Carlo method under different annealing conditions: (a)\SI{240}{\second} at \SI{750}{\celsius}, (b)\SI{60}{\second} at \SI{800}{\celsius}, (c)\SI{180}{\second} at \SI{800}{\celsius}, (d)\SI{10}{\second} at \SI{900}{\celsius}.\cite{Koumetz2003a, Ketata1999, Marcon2003}} \label{fig:kmc result}
\end{figure}

Based on the parameters derived from the continuum model, we also reproduced the experiments of grown-in Be diffusion in InGaAs by KMC simulations.
Here, we assume that Be diffusion in InGaAs can be described by the following kick-out mechanism:
\begin{equation}\label{eq:kmc mechanism}
\begin{aligned}
  \mathrm{Be}_{\mathrm{III}}^{++} + &\mathrm{III}_{\mathrm{i}}^0 \rightleftharpoons &\mathrm{Be}_{\mathrm{i}}^{++}, \\
              &\Updownarrow              &\Updownarrow  \\
  \mathrm{Be}_{\mathrm{III}}^{++} + &\mathrm{III}_{\mathrm{i}}^- \rightleftharpoons &\mathrm{Be}_{\mathrm{i}}^+, \\
                  &                           &\Updownarrow \\
                  &                           &\mathrm{Be}_{\mathrm{i}}^0,
\end{aligned}
\end{equation}
where $\mathrm{Be}_{\mathrm{i}}^0$ denotes a neutral Be interstitial.
Using parameters show in \autoref{tab:kmc parameters}, we successfully fit experiment data as shown in \autoref{fig:kmc result}.
\begin{table*}
  \centering
  \caption{Atomistic parameters of the atom species involved in Be diffusion used in the MMonCa. $D_{m, 0}$ is the diffusivity prefactor, $E_m$ the migration energy and $e_t$ the transition level measured from the valence band edge. Prefactors for equilibrium concentrations and formation energies are $[X]_0^{eq}$ and $E_F$. Parameters left blank will be calculated by MMonCa using \autoref{eq:reaction rate} and \autoref{eq:charge states transformation}.} \label{tab:kmc parameters}
  \begin{tabular*}{\textwidth}{c| @{\extracolsep{\fill}} cccccc}
    \hline
    \hline
       & $\mathrm{Be}_{\mathrm{III}}^{++}$ & $\mathrm{III}_{\mathrm{i}}^0$ & $\mathrm{III}_{\mathrm{i}}^-$ & $\mathrm{Be}_{\mathrm{i}}^0$ & $\mathrm{Be}_{\mathrm{i}}^+$ & $\mathrm{Be}_{\mathrm{i}}^{++}$ \\
    \hline
    ${D_{m,0}\times\si{(\square\cm\per\second)}}$ & 0 & \num{8.40e-3} & \num{8.40e-3} & 0 & \num{5.34e3} & \num{6.71e-3}\\
    $E_m(eV)$ & 5 & 1.44 & 1.44 & 5 & 3.09 & 1.41\\
    $e_t(T=0)$(eV) &   &   & 0.55 &   & 1.00 & 0.80\\
    $[X]_0^{eq}(\times\SI{e25}{\cm^{-3}})$ & 1 & \num{5.00e25} &   & \num{1.33e4} &   &  \\
    $E_F(eV)$ & 0 & 2.13 &  & 1.48 &   &  \\
    \hline
    \hline
  \end{tabular*}
\end{table*}

\par
We notice in \autoref{tab:kmc parameters} that while the charge states in our KMC simulations are consistent with the DFT results, the migration energies are quite different.
This is not surprising since we had to adopt several simplifications in the KMC simulation imposed by the program; specifically, the simulation neglected reactions involving As and treated Ga and In as the same object.
In principle, Ga and In need to be treated differently in the KMC simulations because their reaction parameters are different.
Also, reactions involving As have sufficiently small reaction energies and barriers to contribute substantially to the process (see  \autoref{fig:BeAs kickout path} and \autoref{tab:kick-out mechanism}).
Therefore, the KMC parameters are not directly comparable with the DFT results and one needs instead to adopt a more advanced KMC simulation in the future.
The migration energies of $\mathrm{Be}_{\mathrm{i}}^+$ and $\mathrm{Be}_{\mathrm{i}}^{++}$ are quite different, which is consistent with the continuum method.

\section{Summary} \label{sec:level4}
\par
In this work, a physically motivated multiscale modeling of grow-in Be diffusion in InGaAs has been presented.
In order to evaluate the importance of different diffusion mechanisms, we investigated the reaction energies and elementary diffusion processes in Be-doped \ce{In_{0.5}Ga_{0.5}As} using DFT.
The energy barrier for Frank-Turnbull mechanism was found to be much higher than for the kick-out mechanism, therefore, the kick-out mechanism is likely to be primarily responsible for Be diffusion in the experimental temperature range.
In contrast to previous models, the roles of Ga and In are found to be different in the kick-out reactions, specifically, kicking out of Ga is an exothermic reaction while kicking out of In is endothermic.
Therefore, accurate simulations should treat Ga and In separately.
The reaction energy for kicking out of As is comparable to that for kicking out of Ga and In, thus one should revisit the commonly accepted assumption that As is not involved in Be diffusion.
Besides providing a mechanistic insight, the ab initio simulations also provided physically motivated parameters which served as input for the continuum and KMC simulations.
Continuum modeling indicated that different charge states of Ga and Be interstitials are contributing to the diffusion mechanism at different temperatures and both continuum and KMC simulations agree well with experimental results.
The differences in optimal KMC parameters and DFT suggested parameters points to the deficiency of common approximate treatments of Ga, In and As.

\section*{Acknowledgement}
We acknowledge industry financial support provided by GlobalFoundries, and a research scholarship provided by the Energy Research Institute at Nanyang Technological University (ERIAN).

\bibliographystyle{apsrev4-1}
\bibliography{InGaAs_paper}

\end{document}